\documentclass[aps,prl,superscriptaddress,twocolumn]{revtex4}
\usepackage[dvips]{graphicx}
\usepackage{bm}

\def\Tspin{\ensuremath{T_\mathrm{S}}}
\def\Tmelt{\ensuremath{T_\mathrm{M}}}
\def\Tsoft{\ensuremath{T_\mathrm{soft}}}
\def\Tss{\ensuremath{T_\mathrm{ss}}}

\def\Rvec{\bm{R}}
\def\qvec{\bm{q}}
\def\fvec{\bm{f}}
\def\uvec{\bm{u}}
\def\beq{\begin{equation}}
\def\eeq{\end{equation}}

\begin{document}
%%%%%%%%%%%%%%%%%%%%%%%%%%%%%%%%%%%%%%%%%%%%%%%%%%%%%%%%%%%%%%%%%%%%%%%
%======================================================================
% Per il titolo: loro ci vorrebbero far togliere la parola nanofriction
%======================================================================
\title{Peak Effect versus Skating in High Temperature Nanofriction}
\author{T. Zykova-Timan} 
\altaffiliation[Present address: ]{Computational Science Department of
Chemistry and Applied Biosciences, ETH Zurich USI-Campus, Via Giuseppe
Buffi 13, LUI CH-6900 Lugano, Switzerland}
\affiliation{Scuola Internazionale Superiore di Studi Avanzati
(SISSA) and DEMOCRITOS, Via Beirut 2-4, I-34014 Trieste, Italy}
\author{D. Ceresoli}
\affiliation{Scuola Internazionale Superiore di Studi Avanzati
(SISSA) and DEMOCRITOS, Via Beirut 2-4, I-34014 Trieste, Italy}
\author{E. Tosatti}
\affiliation{Scuola Internazionale Superiore di Studi Avanzati
(SISSA) and DEMOCRITOS, Via Beirut 2-4, I-34014 Trieste, Italy}
\affiliation{International Centre for Theoretical Physics (ICTP),
P.O.Box 586, I-34014 Trieste, Italy}
\date{\today}

%%%%%%%%%%%%%%%%%%%%%%%%%%%%%%%%%%%%%%%%%%%%%%%%%%%%%%%%%%%%%%%%%%%%%%%
\begin{abstract}
%%%%%%%%%%%%%%%%%%%%%%%%%%%%%%%%%%%%%%%%%%%%%%%%%%%%%%%%%%%%%%%%%%%%%%%
%======================================================================
% Abstract: lo vogliono corto, un digest. L'ultima frase deve sottolineare
% le implicazioni pratiche o la proposta di un esperimento
%======================================================================
The physics of sliding nanofriction at high temperature near the substrate 
melting point $\Tmelt$ is so far unexplored. We conducted simulations of hard
tips
sliding on a prototype non-melting surface, NaCl(100), revealing in this regime
two distinct and opposite phenomena for plowing and for grazing friction. 
We found a frictional drop close to $\Tmelt$ for deep plowing
and wear, but on the contrary a frictional rise for grazing, wearless sliding.
For both phenomena we obtain a fresh microscopic understanding, relating the
former to ``skating'' through a local liquid cloud, the latter to linear
response properties of the free substrate surface. It is argued that both
phenomena should be pursued experimentally, and much more general than the
specific NaCl surface case. Most metals in particular possessing one or more
close packed non-melting surface, such as Pb, Al or Au(111), that should
behave quite similarly. 

%Plowing friction of a sharp surface-indented tip shows a drop
%of the frictional force close to $\Tmelt$. Here the tip, accompanied
%by a small moving liquid cloud, effectively ``skates'' through the hot
%solid. At the opposite extreme, the zero load grazing friction of a blunt
%flat tip, initially free sliding at low temperature, shows a sharp surge
%as the surface lattice, while still solid, turns increasingly soft and
%compliant near $\Tmelt$. Linear response theory suggests that the surge
%of grazing friction could be seen as an analogue of the celebrated peak
%effect of sliding vortices in the mixed state of type II superconductors.

%%%%%%%%%%%%%%%%%%%%%%%%%%%%%%%%%%%%%%%%%%%%%%%%%%%%%%%%%%%%%%%%%%%%%%%
\end{abstract} 
\maketitle
%%%%%%%%%%%%%%%%%%%%%%%%%%%%%%%%%%%%%%%%%%%%%%%%%%%%%%%%%%%%%%%%%%%%%%%

Thanks to the advent of increasingly sophisticated tip-based tools,
the physics of nanofriction is now a field of growing importance. There
is in particular a need for a deeper understanding of the physics of
friction in extreme regimes including very high temperatures (HT) where
the sliding materials approach their melting points. In everyday life,
high frictional temperatures are routinely attained in a variety of
situations including machining, motors, braking, etc~\cite{iavf}. While
rather complex phenomena are generally involved in many such practical
situations, at the fundamental level even the very basic microscopic
friction processes taking place close to the substrate melting point
are not sufficiently explored, characterized and understood. For bare
metals on metals, for example, a high speed \emph{drop} of wear-related
friction is generally reported~\cite{bowden-tabor}, and reasonably attributed
to softening caused by high (``flash'') temperatures (``if one could move fast
enough, one could ski on copper mountains\dots''~\cite{bowden-tabor}), but
not yet pursued with surface science standards. Conversely, dry grazing
friction between hard materials has been known to \emph{rise} at HT,
ostensibly due to the onset of some diffusion~\cite{bowden-tabor};
again however that has not been further followed up microscopically. At
the present stage, it is just not clear when, how and why closeness to
the substrate melting point will imply a drop or a rise of frictional
forces under controlled contact circumstances. Nanofriction presently
offers a fresh opportunity to address the microscopic physical facts
behind these phenomena which are receiving increasing
attention.~\cite{krylov05}
Recent simulation studies have addressed metallic friction between a 
hard indenter and a soft surface, at high velocities and at tip-driven
local temperatures near and above the bulk melting point~\cite{hammerberg04}.
Also the simulated motion of a rigid tip across a Si surface at room
temperature 
showed an interesting tip and temperature driven local phase transition 
leading to a reduced friction~\cite{zhang04}.

\begin{figure}
\begin{center}
  \includegraphics[width=0.95\columnwidth]{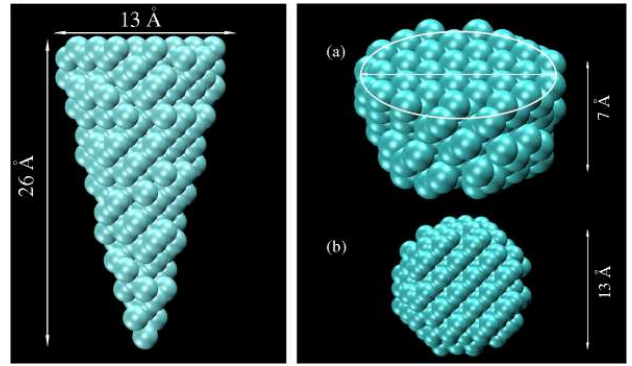}
  \caption{Left panel: rigid diamond tip constructed for plowing
  friction.  The tip length is $\sim$~26~\AA, the diameter is about
  13~\AA. Right panel: rigid diamond tip constructed for wearless
  friction. (a) side view; (b) top view. The tip length is $\sim$~7~\AA,
  the diameter is about 13~\AA.}
  \label{fig:tips}
\end{center}
\end{figure}

Considering that most AFM nanotips slide normally at very low speed, and 
can hardly perturb the massive substrate, our attention is focused on 
friction under conditions of essential closeness to thermal equilibrium, 
however at temperatures arbitrarily near or even marginally above the 
substrate $\Tmelt$. We arrive at a first characterization of possible 
scenarios by means of simulation and theory. Our starting point are 
molecular dynamics (MD) simulations of sliding nanotips on the hard 
crystalline surface of a solid close to its bulk melting point $\Tmelt$. 
We actually recover both the high
temperature friction drop for deep plowing and wear, and the frictional
rise for weak load wearless sliding; and for both regimes we obtain a
fresh microscopic understanding.

For a meaningful case study we require first of all a careful choice of
surfaces and tips. In fact, close to $\Tmelt$ most solid surfaces wet 
themselves -- in full equilibrium conditions and without any perturbing 
tips -- with a microscopically thin film of melt (surface melting)
~\cite{report}.  The liquid film will generally
``jump to contact'' with the tip long before closest approach -- to some
extent ruining the experiment, as indeed observed~\cite{kuipers93}
and predicted~\cite{tomagnini93} long ago. This unwanted complication
could be avoided for example by resorting to tips that are not wetted by
the substrate. Alternatively one could, as we do in the present work,
choose a hard, \emph{non-melting} crystalline substrate facet exhibiting
no spontaneous surface melting at all -- a surface that will spontaneously
remain solid and crystalline up to $\Tmelt$~\cite{report}.  One added
advantage of this choice is the possibility to distinguish between very
different frictional regimes, such as deep plowing (involving plastic flow and
heavy wear)
and superficial (wearless) grazing friction, a distinction otherwise
made impossible by surface melting. We did not consider inhomogeneous
surfaces, simply kept from melting by some kind of protective coating,
for they will be very difficult to control. Close-packed metal surfaces,
such as Pb, Al or Au(111) are known to be nonmelting~\cite{report}
and would serve the purpose. Eventually, the very stable alkali halide
NaCl(100) surface proved close to an ideal choice because of its extreme
and recently understood nonmelting habit~\cite{PRL05}. One added bonus
of alkali halide surfaces is that they are already choice substrates for
abundant experimental nanofriction work at room temperature~\cite{gnecco}.

The high temperature behavior of NaCl was simulated by the classic
Born-Mayer-Huggins-Fumi-Tosi (BMHFT) two body potential~\cite{fumi}. Bulk
molecular dynamics (MD) simulations~\cite{PRL05,JCP}, as well as
thermodynamic integration~\cite{frenkel} yield for that model a bulk
melting temperature $\Tmelt=$1066~K (experimental value 1074~K). The
NaCl(100) surface was studied using periodically repeated slabs
consisting of 12~$\div$~24 planes of 100 NaCl units each separated by
100~$\div$~120~\AA\ of vacuum. Computational details were similar to
those of Ref.~\cite{JCP} The perfect, tip free NaCl(100) was found to
remain indefinitely solid and free of diffusion at all temperatures
up to $\Tmelt$ and, in a metastable state, even above $\Tmelt$ up
to a ``surface spinodal temperature'' $\Tspin \sim$~1200~K, for our
longest simulation times $\sim$~200~ps. This qualifies NaCl(100) as a
nonmelting surface, in agreement with its very poor wetting by molten
NaCl~\cite{mutaftschiev,report,PRL05,JCP}.

To simulate nanofriction we modeled a rigid diamond tip apex, for
which deformations are negligibly small. For indentation and deep wear
(plowing) simulations we used the sharp $\sim$~400 atom conical tip
(Fig.~\ref{fig:tips}a). For wearless (grazing) frictional simulations
we used instead a blunt flat tip whose contact plane is composed
of  $\sim$~200 atoms (Fig.~\ref{fig:tips}b) forming a diamond (111) plane
of $\sim$~13~\AA\ diameter.  We slid our tips at constant velocities
and purposely avoided including any cantilevers with their mechanical
deformations, and related stick-slip phenomena. Thanks to the relatively
minor perturbation represented by the sliding tip, even in the plowing
case no additional heat sink other than temperature control through standard
velocity rescaling was necessary to ensure efficient dissipation
of the Joule heat.

\begin{figure}
\begin{center}
  \includegraphics[angle=270,width=0.95\columnwidth]{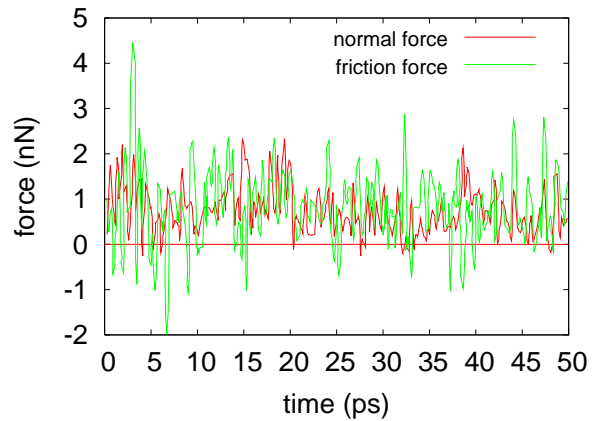}
  \caption{Instantaneous forces in a plowing friction
  simulation of NaCl(100) under constant penetration of 6~\AA, speed
  of 50~m/s, and $T=$~300~K}
  \label{fig:forces}
\end{center}
\end{figure}

We address first of all plowing friction, with the aim of understanding
wear forces and their dependence on temperature. For that, we indented
the surface with the sharp tip and slid it at constant velocity $v$
(generally between 25 and 50 m/s) parallel to the NaCl(100). Because
under constant load at high temperatures the tip penetration will
generally be unstable with time, we worked in a constant penetration
mode, the tip wedge entering the NaCl(100) surface by a fixed depth
$d=$~6~\AA.  The forces acting on the tip were collected during the
simulated plowing along the $x$ direction -- parallel to the surface --
across the 60~$\times$~60~\AA~$^2$ wide surface.  

\begin{figure}
\begin{center}
  \includegraphics[angle=270,width=0.95\columnwidth]{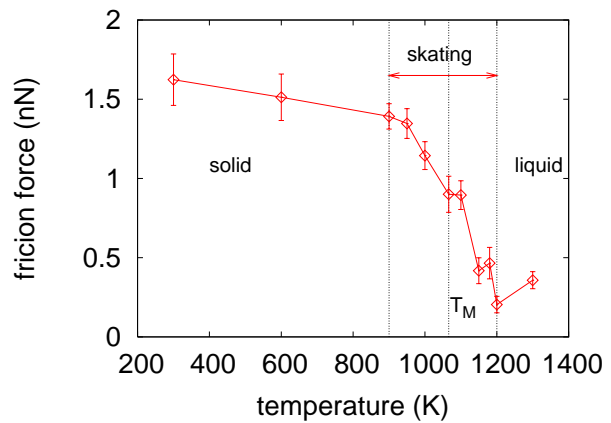}
  \caption{Averaged plowing frictional force $\langle F_x \rangle$ as a
  function of temperature. Note the drop at $T_{ss} \sim T_m -$~150~K. The
  frictional stress $\sigma =\langle F_x \rangle / A$ can be obtained using
  an active tip plowing area of $\sim$6~\AA$^2$. }
  \label{fig:friction}
\end{center}
\end{figure}

The vertical and in plane forces $F_z(t)$ and $F_x(t)$ are shown in
Fig.~\ref{fig:forces}. The large noise reflects the discontinuous nature
of wear, including plastic flow and the formation and displacement of debris.
As a function
of temperature, the mean frictional force $\langle F_x(t)\rangle$,
shown in Fig.~\ref{fig:friction} clearly shows two regimes.

Up to $ T= \Tsoft \sim$~900~K~$\simeq$~0.85~$\Tmelt$, the plowing frictional
stress is approximately constant, corresponding to a friction coefficient
$\mu =\langle F_{x}/F_{z} \rangle \sim$~1. This large value is
naturally attributed to plastic deformation, breaking and 
reforming of atomic bonds, and the digging of a trough. For macroscopic
indenters onto larger size substrates, plowing would in addition generally
cause dislocations to appear.    
Our nanosized, poorly indented tip is in fact unable to generate dislocations
and only leads to local plastic deformation and eventually local melting
very near $\Tmelt$.
The typical magnitude of $F_x(t)$ is around 1 $nN$, which given
the small tip active area is equivalent to a stress of about 6 GPa, a value 
close to the estimated yield stress of 8.3 GPa of NaCl against (100) shear
near $\Tmelt$. 

\begin{figure}
\begin{center}
  \includegraphics[width=0.95\columnwidth]{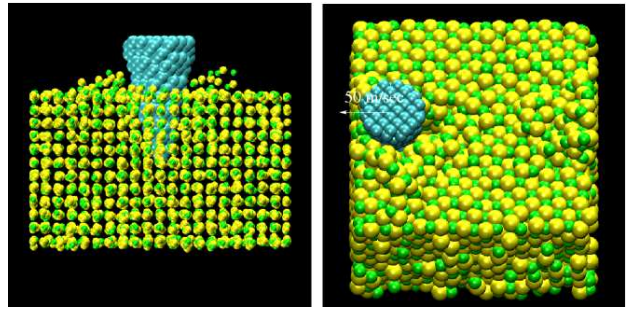}
  \caption{At \Tmelt\ a hard tip plowing through solid NaCl(100) is
  surrounded by a local liquid cloud that moves with it. Light blue
  spheres are tip atoms, green stand for Na$^{+}$ and yellow stand for
  Cl$^{-}$ respectively. Note the fast healing furrow behind the tip.
  For clarity of illustration, the penetration here is 18~\AA.}
  \label{fig:tip_cloud}
\end{center}
\end{figure}

An approximate frictional energy balance at 300~K shows that about 20\%
is spent to plastically deform and scratch the material, and the remaining 80\% 
is dissipated as
heat, similar to typical figures given in the literature~\cite{persson}.
The work-induced local melting of NaCl(100) near the tip occurs at essentially
constant temperature, the mechanical energy transfered into heat of melting, 
plus phonons
that are effectively absorbed by the temperature control through velocity
rescaling.
Above $\Tsoft \sim \Tmelt -$~150 ~K, the frictional force drops and 
remains low across the $\Tmelt$ and above up
to $\Tss \sim$~1200~K where the surface spontaneously melts.  The drop
of friction above $\Tsoft$ resembles the experimental one reported
for metals on metals, and is reminiscent of that observed for hard sliders 
on ice. In ice~\cite{persson}, the accepted view is that the low skating
friction
is due to a thin sliding-induced  lubricating liquid layer.  In our
simulation, we find that indeed the frictional drop is accompanied
by a thin local cloud of displaced Na and Cl ions which surround
the tip with liquid-like mobilities (see Fig.~\ref{fig:tip_cloud}).
The local melting is motion-induced, and not due \emph{e.g.}, to uniaxial
stress-induced lowering of the melting point.~\cite{levitas06,tartaglino03}
Coated by this nearly liquid shroud (Fig.~\ref{fig:tip_cloud}) the tip
effectively skates. The ease with which the tip wades through the solid
is remarkable. The scratch produced by the moving tip, permanent at low
temperatures, heals rapidly away at higher temperatures, (see movie in
the supplementary material) with a characteristic time which is in the 
ps range at $\Tmelt$. The fast resolidification behind the tip restores
the equilibrium solid nature of the NaCl surface at \Tmelt~\cite{PRL05,JCP},
and the resulting latent heat is dissipated away as phonons.

\begin{figure}
\begin{center}
  \includegraphics[angle=270,width=0.9\columnwidth]{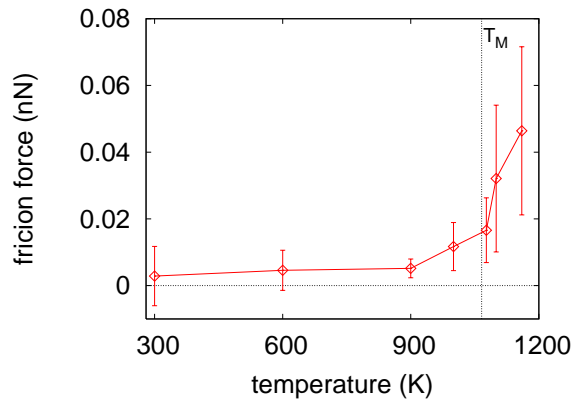}
  \caption{Frictional force of a grazing tip on NaCl(100). Note the
  very low friction, sharply increasing at $\Tmelt$ and above. Here the
  effective contact surface area was $A \sim$170~\AA$^2$.}
  \label{fig:peak}
\end{center}
\end{figure}

Totally different from plowing friction are the results, shown in Fig.%
~\ref{fig:peak}, for wearless friction of a large flat tip, grazing
the NaCl(100) surface under zero load. Here the tip contacts the substrate by
mere adhesion, the adhesive energy $V_\mathrm{ad}\simeq$~0.7~eV equivalent to
an effective load $F_{z} \sim -V_\mathrm{ad}/h$ where $h$ is of order of
half the interplanar spacing (all results remain very similar for small
positive loads). At low temperatures the strongly fluctuating frictional
force $F_{x}(t)$ (not shown) averages to a very small value, yielding a
tiny friction coefficient $\mu = \langle F_{x}\rangle/F_{z}\sim$~0.007
at 300~K. Such nearly frictionless sliding of two hard incommensurate
crystal lattices is not surprising, and resembles that recently reported
for a rotated graphite flake on graphite~\cite{frenken}. As temperature
grows, an initially moderate thermal growth of $\langle F_{x}\rangle$
up to 900~K gives way to an important nonlinear frictional surge near
$\Tmelt$. This high temperature frictional peak makes a sharp contrast
with the high temperature drop just seen in plowing friction on the
same surface. Here the grazing frictional energy is dissipated mostly
through generation of phonons in the solid substrate, plus some modest
tip-induced drift of diffusive surface atoms taking place at the highest
temperatures. Unlike low temperature plowing, the solid surface is left,
after dissipating away the small amount of frictional heat, very much
in the same state it was before friction. This observation justifies
an attempt at treating grazing friction within linear response theory,
assuming a steady sliding state, and negligible influence of the tip on
the surface.

Modeled as a hypothetical impurity-like particle, assumed free
to move along $(xy)$ on the hot surface, a tip will experience random
forces -- Brownian kicks. In the absence of an external force, it would
undergo 2D diffusion; under a planar force $\fvec$, it should drift 
with average velocity $v$ proportional to $\fvec$. 
Simulations at 900~K with increasing velocities between 15 and 50~m/s confirmed
an approximately linear increase of friction force. Here the viscous behavior
of grazing friction is due to random kicks from large thermal vibrations,
rather than to viscoelastic properties of the hot surface, still perfectly
solid and crystalline.  
Assuming a Langevin equation $M \ddot{\Rvec} =
-\eta M \dot{\Rvec} + \fvec(t)$ for the in-plane Brownian motion of the
idealized (force-free) tip centered at  $\Rvec$, the zero frequency self
correlation of the random force $\fvec(t)$ is directly related to friction, in
the simple form~\cite{hansenbook}
\beq
  \eta = \frac{1}{3M k_{B} T} \int_{0}^{\infty} dt \langle \fvec(t)
  \fvec(0)\rangle
\eeq 
where it is moreover assumed, as usual that \mbox{$\langle \fvec(t)
\rangle = 0$}.  Within linear response, and for
negligible tip influence over the surface, the random force acting on
the tip can be taken to be proportional to the surface atom displacements
$\uvec_{i}(t)$.  Following Ying \emph{et al.}~\cite{granato,ying90} it is
convenient to use 2D $k$-space $\uvec_i(t) = \int d^2q\, \uvec_{\qvec}(t)
e^{-i\qvec\cdot\Rvec_i}$ and write the random tip force in the form
$\fvec(t) = \sum_{\qvec} C_{\qvec}\uvec_{\qvec}(t)$ where $C_{\qvec}$
is a coupling function expressing the assumed proportionality between
kicks onto the tip with individual displacements  $\uvec_i(t)$ and the
random force $\fvec(t)$ felt by the tip.  Inserting this expression into
the equation for $\eta$, we obtain
\beq
  \eta = \frac{1}{3 M k_{B} T}\sum_{\qvec}|C_{\qvec}|^{2}
  S(\qvec,\omega=0) \label{eq:star}
\eeq
where $S(\qvec,\omega=0) = \int_{0}^{\infty} dt
\uvec^{\star}_{-\qvec}(t) \uvec_{\qvec}(0)$ is the two dimensional
dynamical structure factor of the tip-free surface substrate, proportional
to the substrate atom displacement auto-correlation functions. In
this manner, the grazing friction $\eta$ is reduced to a property of
the unperturbed surface. In the simple present form, it could also be
arrived at directly by observing that the dissipation rate takes in linear
response theory a Fermi Golden rule form, $|C_{\qvec}|^2$ measuring the
perturbation strength, and the dissipative part of the system's linear
response function giving the density of final states. Considering that the
frequencies that are relevant for dissipation (long wavelength phonons
and diffusional processes) are extremely small compared to e.g. the
solid substrate Debye frequency, one can set  $\omega=$~0 recovering
(disregarding factors) eq.~(\ref{eq:star}) above.

\begin{figure}
\begin{center}
  \includegraphics[angle=270,width=0.9\columnwidth]{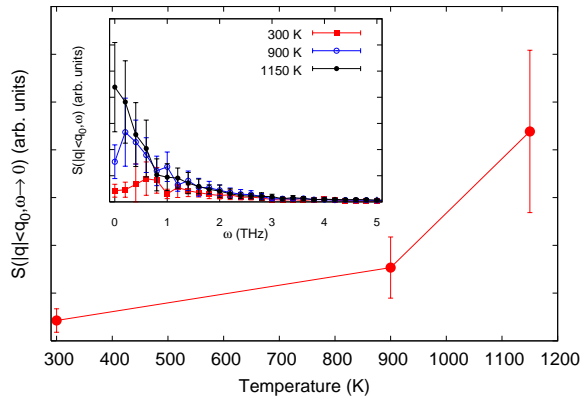}
  \caption{Dynamical surface atom structure factor of NaCl
  $\overline{S}(\omega\rightarrow 0)$ as a function of temperature for
  surface atoms only. Inset, $\overline{S}(\omega)$ for temperature
  from 300~K to 1150~K.}
  \label{fig:s_factor}
\end{center}
\end{figure}

We can now attempt a comparison of the temperature dependence predicted
by linear response theory eq.~(\ref{eq:star}) of grazing friction based
on the dynamical properties of the \emph{tip-free} surface with the
actual simulated friction. Here we deliberately oversimplify by assuming a
T independent coupling $C_{\qvec}$ -- in reality, hard core interatomic
potentials will generally imply a growing value with temperature, since
vertical displacements, which eq.~(\ref{eq:star}) disregards, will be
more important at higher temperature. In $q$ space,  $C_{\qvec}$ will be
nonzero for $q < q_0 \equiv 2\pi/L$, where $L$ is the lateral dimension of 
the tip ($\sim$~13~\AA) or roughly $q_0 \simeq$ 0.5~\AA$^{-1}$.  Since $q_0$
is small, we can simply assume $C_{\qvec} = C_0\,\Theta(q_0 - q)$, whence
\begin{equation}
  \eta = \frac{1}{3 M k_{B} T}|C_0|^2 \int_{q<q_0} d\qvec\, S(\qvec,\omega=0)
\end{equation} 
To check the theory against frictional simulation,
we extracted displacement auto-correlation functions of NaCl(100)
from tip-free surface simulations.  Fig.~\ref{fig:s_factor} shows
$\overline{S}(\omega)=\int_{q<q_0} d\qvec\, S(\qvec,\omega)$ as a function
of temperature, of surface atoms only.
Comparison with Fig.~\ref{fig:peak} is qualitatively
good, although as anticipated some increase of $C_{\qvec}$ would be needed to 
make the agreement quantitative at high temperature.  The sharp rise close
to $\Tmelt$ and $\omega \to 0$ is attributable to large and eventually
(close to $\Tss$) catastrophic softening of the very anharmonic surface
lattice, still solid but extremely compliant in this regime.

A well known analogue of this phenomenon is actually realized in the
physics of type II superconductors close to $H_\mathrm{c2}$. Here the
flux lattice -- whose frictional depinning from the ion lattice and
impurities determines the critical current -- turns soft just before
eventually disappearing~\cite{chaikin}.  This soft state is so compliant
that the pinning force, and the critical current with it, develops a
last sharp peak before dropping to zero at $H_\mathrm{c2}$, a phenomenon
that has been described by very similar formulas to those of sliding
friction~\cite{granato,ying90}. It is thus reasonable to propose that
the high temperature friction increase of a hard grazing slider we just
described is the direct frictional analogue of this \emph{peak effect} in type
II superconductors.

In summary, the sliding of hard tips onto hard nonmelting surfaces
(here modeled by NaCl(100)) reveals new phenomena in high temperature
nanofriction.  As the melting point is approached, friction may either
drop as in skating, or rise as in flux lattice depinning. We believe
that these phenomena should be more general than the specific NaCl
surface context in which they emerged. In particular, most metals
possess at least one close packed non-melting surface, such as Pb, Al
or Au(111)~\cite{report}, that should behave quite similarly to our
results on NaCl(100). High temperature experiments on NaCl itself
should be feasible. Because of the high vapor pressure (0.34 mmHg
at $\Tmelt$), and of a correspondingly high rate of evaporation (we
estimate 1.7~$\times$~10~$^{-5}$~s$^{-1}$, for terraces 50~\AA\ wide), 
the step flow velocity at sublimation conditions at $\Tmelt$ will be
about 10~$^{-5}$~m/s. So long as the tip velocity is larger than this
value, our description of nanofriction is fully applicable. Moreover,
irrespective of sublimation the \emph{terraces} between the steps will
still be well represented by the flat, dry, stable solid surface described
in our simulations. Thus the effects described will be readily observed,
for example once the sliding tip velocity $v$ is faster than the step
flow velocity.

\noindent
\textbf{Acknowledgements:} This work was sponsored by MIUR FIRB RBAU017S8 R004, 
FIRB RBAU01LX5H, and MIUR COFIN 2003 and 2004, as well as by INFM (Iniziativa
trasversale calcolo parallelo). We acknowledge illuminating discussions with,
and much help from, U. Tartaglino.

%%%%%%%%%%%%%%%%%%%%%%%%%%%%%%%%%%%%%%%%%%%%%%%%%%%%%%%%%%%%%%%%%%%%%%%

%%%%%%%%%%%%%%%%%%%%%%%%%%%%%%%%%%%%%%%%%%%%%%%%%%%%%%%%%%%%%%%%%%%%%%%

%\noindent
%\textbf{Corresponding author:} E. Tosatti, \textsl{tosatti@sissa.it}

%\noindent
%\textbf{Acknowledgements:} This work was sponsored by MIUR FIRB RBAU017S8 R004, 
%FIRB RBAU01LX5H, and MIUR COFIN 2003 and 2004, as well as by INFM (Iniziativa
%trasversale calcolo parallelo). We acknowledge illuminating discussions with,
%and much help from, U. Tartaglino.

%%%%%%%%%%%%%%%%%%%%%%%%%%%%%%%%%%%%%%%%%%%%%%%%%%%%%%%%%%%%%%%%%%%%%%%
\end{document}